\documentclass[aps,prb,twocolumn,showpacs,superscriptaddress,10pt]{revtex4-1}\begin{tiny}\end{tiny}
\usepackage{longtable}
\usepackage{amsfonts}
\usepackage[dvipdfmx]{graphicx}
\usepackage{rotating}
\usepackage{hyperref}
\usepackage{dcolumn}
\newcolumntype{d}{D{.}{.}{3.5}}

\begin{document}

\title{Magnetic gap opening in rhombohedral-stacked multilayer
  graphene from first principles}

\author{Bet\"{u}l Pamuk}
 \affiliation{CNRS, UMR 7590 and Sorbonne Universit\'{e}s, UPMC Univ Paris 06, IMPMC - Institut de Min\'{e}ralogie, de Physique des Mat\'{e}riaux, et de Cosmochimie, 4 place Jussieu, F-75005, Paris, France}
 
\author{Jacopo Baima}
  \affiliation{CNRS, UMR 7590 and Sorbonne Universit\'{e}s, UPMC Univ Paris 06, IMPMC - Institut de Min\'{e}ralogie, de Physique des Mat\'{e}riaux, et de Cosmochimie, 4 place Jussieu, F-75005, Paris, France}

\author{Francesco Mauri}
 \email{francesco.mauri@uniroma1.it}
 \affiliation{Dipartimento di Fisica, Universit\`{a} di Roma La Sapienza,
Piazzale Aldo Moro 5, I-00185 Roma, Italy}
 \affiliation{Graphene Labs, Fondazione Istituto Italiano di Tecnologia, Via Morego, I-16163 Genova, Italy}
  
\author{Matteo Calandra}
   \email{matteo.calandra@impmc.upmc.fr}
   \affiliation{CNRS, UMR 7590 and Sorbonne Universit\'{e}s, UPMC Univ Paris 06, IMPMC - Institut de Min\'{e}ralogie, de Physique des Mat\'{e}riaux, et de Cosmochimie, 4 place Jussieu, F-75005, Paris, France}
   
\date{\today}


\begin{abstract}

We investigate the occurrence of magnetic and charge density wave
instabilities in rhombohedral-stacked multilayer (three to eight layers) graphene 
by first principles calculations including exact exchange.
Neglecting spin-polarization, an extremely flat surface band centered
at the special point ${\bf K}$ of the Brillouin zone occurs at the
Fermi level. 
Spin polarization opens a gap in the surface state by
stabilizing an antiferromagnetic state. 
The top and the bottom surface layers are weakly ferrimagnetic in-plane
(net magnetization smaller than $10^{-3}\mu_B$),
and are antiferromagnetic coupled to each other.
This coupling is propagated by 
the out-of-plane antiferromagnetic coupling between the nearest neighbors.
The gap is very small in a spin-polarized generalized gradient
approximation, while it is proportional to the amount of exact
exchange in hybrid functionals. 
For trilayer rhombohedral graphene it is $38.6$ meV in PBE0, 
in agreement with the $42$ meV gap found in experiments. 
We study the temperature
and doping dependence of the magnetic gap. 
At electron doping of $n \sim 7 \times 10^{11}$ cm$^{-2}$ the gap closes. 
Charge density wave instabilities with $\sqrt{3}\times\sqrt{3}$
periodicity do not occur.

\end{abstract}

\maketitle

\begin{figure}[!thb]
        \centering
                \includegraphics[clip=true, trim=0mm 0mm -10mm -10mm, width=0.48\textwidth]{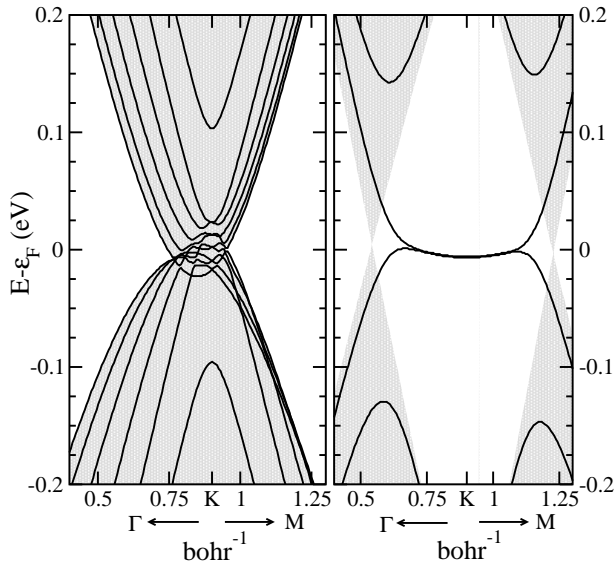}
        \caption{Electronic structure (black lines) of a 10-layer
          Bernal (left) and rhombohedral (right) -stacked multilayer
          graphene calculated with the LDA functional.
          The gray regions are the bulk electronic
          structures projected over the surface.}
        \label{fig:BulkoverS}
\end{figure}

In a solid, at low enough density, the Coulomb energy dominates the
single particle energy and
electronic instabilities 
such as magnetic phases or even Wigner crystallization become possible. 
A reduction of the single-particle energy, 
and a consequent enhancement of the electron-electron interaction,
can be obtained by considering a  metallic system with a very flat
single-particle band-dispersion. 
This unfortunately does not happen in
graphene where the high Fermi velocity prevents electronic instability.
The situation is different, however, in weakly doped Bernal-stacked (AB)
even-multilayer graphene
(see Fig. \ref{fig:BulkoverS}), 
as the single-particle bands become massive.
Indeed, a spontaneously gapped 
ground state is already observed in suspended bilayer
\cite{Schonenberger2012, MacDonald2012} 
and four-layer graphene (1.5 meV gap)\cite{Morpurgo2015}.

Even more favorable to electronic
instabilities is rhombohedral-stacked (ABC) multilayer graphene.
Neglecting spin polarization, tight-binding
and density functional theory (DFT)
calculations find bulk rhombohedral
graphite to be metallic with a high Fermi velocity
(see grey region in Fig.  \ref{fig:BulkoverS}). 
On the contrary, within these approximations,
rhombohedral-stacked multilayer graphene (RSMG) displays the occurrence of an extremely flat
surface state at the Fermi level (see Fig.  \ref{fig:BulkoverS}) located
at the ${\bf K}$ point of the graphene Brillouin zone (BZ)
 \cite{Koshino2013,MacDonald2013,Henrard2006,Taut2011,Wu2011,Otani2010}. 
The extension of the surface state in the BZ increases with increasing thickness and
saturates at $\approx 7$ layers.
Its bandwidth is at most $2$ meV for flakes of fewer than eight layers.
The extremely reduced bandwidth makes RSMG
one of the strongest correlated systems known nowadays and an
ideal candidate for correlated states even in the absence of $d$
orbitals.

On the experimental side, only recently has it been possible to synthesize
RSMG. Ouerghi {\it et al.} \cite{Thomas2015} found five-layer RSMG on top of a cubic SiC
substrate to be metallic with a very high density of states at the
Fermi level, in agreement with spinless DFT predictions. More recent
work on suspended ABC trilayer \cite{Guinea2014} 
seems to suggest that a gap as large as $42$ meV occurs
at the Fermi level. However, in transport measurements with singly gated devices,
where the charge density and the potential difference cannot be controlled independently,
the gap is much smaller \cite{Lau2011}. 
The origin of this gap state has been suggested to
be related to a magnetic state in which the external layers are
antiferromagnetic coupled while the in-plane spin state is ferrimagnetic
\cite{Guinea2014,Throckmorton2012}. 
Finally, it is worth mentioning that it has been claimed that RSMG flakes composed
of 17 layers can be isolated from 
kish graphite  \cite{Faugeras2016}. However it is still unknown if a
gap opens in these samples.

In this work, by using density functional theory calculations with several
hybrid functionals  we investigate the occurrence of charge and magnetic instabilities in RSMG. 
We also discuss the effect of doping and temperature on the
interaction induced gap.

DFT calculations are performed using the CRYSTAL code \cite{Crystal14} with the 
triple-$\zeta$-polarized Gaussian type basis sets for the C atoms \cite{Cbasis}.
The PBE0 \cite{PBE0} 
and other hybrid functionals with several degrees of exact exchange 
(see Appendix \ref{XCfcnal})
have been used for DFT calculations. 
The band flatness and the extreme localization of the low energy states around the 
special point ${\bf K}$ require an ultradense sampling 
with an electronic k mesh of $516\times516\times1$. We used
real space integration tolerances of 7-7-7-15-30,
and with an energy tolerance of $10^{-11}$ Ha for the total energy convergence.
Fermi-Dirac smearing for the occupation of the electronic states 
is used for all of the calculations.
The density of states is obtained with a Gaussian smearing of 0.00005 Ha.
In the magnetic case, we fix the magnetic state in the first iteration
of the self-consistent cycle and then we release the constraint. 
We choose an in-plane lattice parameter of $a=2.461$ \AA,
and an inter-plane distance of 3.347 \AA. 
A vacuum of 10.04 \AA\, is placed between the periodic images
along the $z$ direction. 

We consider $N\ge3$, $N$ being the number of layers. 
We start the simulation from an initial magnetic state having 
ferromagnetic coupling in the surface layers and antiferromagnetic 
coupling between them. In both spin polarized
PBE \cite{PBE} and exact exchange functionals (HSE06, PBE0, etc.), we
always converge to 
a magnetic state that is
(1) globally, an antiferromagnetic spin state,
namely for each spin up band we have a spin down band degenerate in energy but localized on different atoms;
(2) ferrimagnetic within the surface layers where the two atoms have opposite spins
with slightly different magnetic moment; and
(3) with antiferromagnetic coupling between the top and bottom layers.
The magnitude of the spin drops significantly farther away from the surface.
This magnetic state
is similar to the one found in bilayer and trilayer
graphene in the framework of Hubbard-like models
\cite{Guinea2014,Throckmorton2012} (the so-called layered
antiferromagnetic state); however, here the module of the magnetic
moments in each outer surface layer is identical up to 0.001 $\mu_B$, meaning
that we are much closer to an in-plane antiferromagnetic state
than a ferrimagnetic state.

\begin{figure}[!thb]
        \centering
                \includegraphics[clip=true, trim=0mm 0mm 0mm 0mm, width=0.48\textwidth]{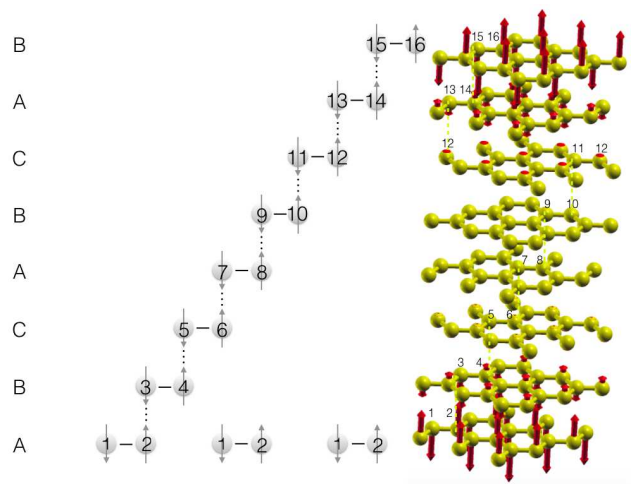}
        \caption{Left: Schematic of the rhombohedral graphene with ABC stacking sequence.
        Solid lines represent the in-plane covalent bond, 
        and dashed lines represent the out-of-plane weak interlayer bond.
        The arrows represent only the direction of the spin,
        as the magnitude is layer dependent.
        Right: The atomic structure of rhombohedral eight-layer graphene.
        }
        \label{fig:struc}
\end{figure}

Figure \ref{fig:struc} shows a schematic diagram of rhombohedral graphene,
with arrows representing the direction of the spin for all layers up to eight layers.
In rhombohedral graphene, the total number of atoms is $N_{\rm atom}=N$,
with each layer having two covalent bonded C atoms,
only one of which is weakly bonded to the neighboring layer.
The labeling of the atoms starts from the bottom layer, 
where atom 1 is covalent bonded to atom 2,
and atom 2 is weakly bonded to atom 3 of the next layer,
and the labeling follows the covalent to weak interlayer bond.
The magnetic order we find is such that the odd-numbered atoms have down spin,
and the even numbered atoms have up spin, 
with the spin magnetic moments such that $\mu_i=-\mu_{{\rm Natom}-i+1}$.
The outermost surface layers are ferrimagnetic with $\mu_1 > \mu_2$.
In-plane and out-of-plane nearest neighbors are antiferromagnetic coupled.

Within spin-polarized PBE, the local magnetic moments are essentially zero.
The increase of the exact exchange component in the functional enhances
the magnetic moments. 
The magnetic moments in the
outermost layers are larger for eight-layer thick flakes and within PBE0 they are as large as
$10^{-2}$ bohr magnetons in magnitude (see Table \ref{table:spins}).
The magnetic moments for other functionals
are shown in Appendix \ref{XCfcnal}.

\begin{table*}[!htb] \footnotesize
\caption{The magnitude of the spin of each atom in $10^{-3}\mu_B$ for rhombohedral $N$-layer graphene.
		The spins are reported up to $\mu_{\rm Natom/2}$, and 
		the spins of the rest of the atoms,
		as well as the direction of the spin can be obtained by $\mu_i=-\mu_{{\rm Natom}-i+1}$.
        This direction can be matched to Fig. \ref{fig:struc}.}
\centering
\begin{ruledtabular}
        \begin{tabular}{c c c c c c c c c}
        N & $\mu_1=-\mu_{{\rm N}_{\rm atom}}$ & $\mu_2$ & $\mu_3$ & $\mu_4$ & $\mu_5$ & $\mu_6$ & $\mu_7$ & $\mu_8$ \\
        \hline
        3 & -5.28 & 4.60 & -2.73 \\
        4 & -7.32 & 6.33 & -3.22 & 3.11 \\
        5 & -8.56 & 7.38 & -3.62 & 3.43 & -2.60 \\
        6 & -9.19 & 7.90 & -3.82 & 3.58 & -2.34 & 2.29 \\
        7 & -9.75 & 8.40 & -4.11 & 3.84 & -2.38 & 2.30 & -1.93 \\
        8 &-10.04 & 8.64 & -4.23 & 3.94 & -2.34 & 2.24 & -1.69 & 1.66 \\
       \hline
        \end{tabular}
        \end{ruledtabular}
\label{table:spins}
\end{table*}

\begin{figure}[!thb]
        \centering
                \includegraphics[clip=true, trim=5mm 5mm -5mm -15mm, width=0.48\textwidth]{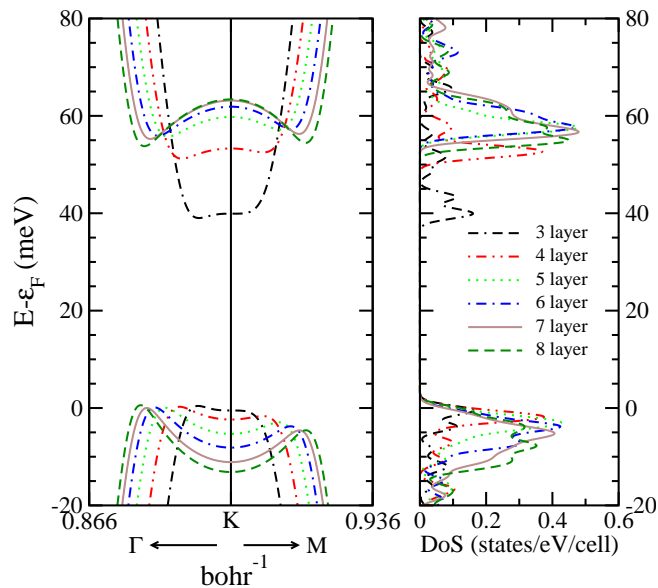}
        \caption{The electronic band structure (left) and density of states (right)
        of the surface states of the magnetic state of the rhombohedral graphene up to eight layers
        calculated with the PBE0 functional.
        The electronic bands are plotted around 0.035 bohr$^{-1}$ of the \textbf{K} point
        along the path $\mathbf{\Gamma} \rightarrow \mathbf{K} \rightarrow \mathbf{M}$,
        and each band is spin degenerate.}
        \label{fig:magneticBands}
\end{figure}

The PBE0 magnetic electronic structure for $3\le N\le8$ is shown in
Fig. \ref{fig:magneticBands}. 
Each band is doubly degenerate with spin up and down states having the same energy.
Therefore globally the system is in an antiferromagnetic state.
This magnetic order results in a gapped
state, with the gap at the {\bf K} point increasing with $N$.
The surface bands are quite flat with a small asymmetry between 
the $\mathbf{\Gamma}\rightarrow \mathbf{K}$ and
$\mathbf{K}\rightarrow\mathbf{M}$ directions
that increases with $N$. 
This behavior is easily understood as for $N\to \infty$ the
surface projected bulk bands of rhombohedral graphite should be
recovered (see Fig. \ref{fig:BulkoverS}).
More insight into the gap opening at ${\bf K}$ is obtained by analyzing the
projected electronic bands on the atoms of the surface layers. 
The surface states are dominated by the $p_z$ orbitals of two C atoms.
On each surface layer, only the atom having no out-of-plane neighboring atoms 
contributes to the surface bands (see Appendix \ref{PBand}).
For comparison, the electronic structure of metallic and paramagnetic state 
is given Appendix \ref{MetalBand}.

We find the band gap for rhombohedral trilayer graphene to be $E_g=38.6$ meV 
with the PBE0 functional. This result is in good agreement with the
experimentally 
reported value of $E_g=42$ meV  in Ref. \cite{Guinea2014}. 
By using other functionals with different percentages of exact exchange and 
range separation we find smaller gaps as shown in Appendix \ref{XCfcnal}.
As the PBE0 gap gives the best agreement with experiment we use
this functional for the rest of the paper.

\begin{figure}[!thb]
        \centering
                \includegraphics[clip=true, trim=0mm 10mm -5mm -15mm, width=0.48\textwidth]{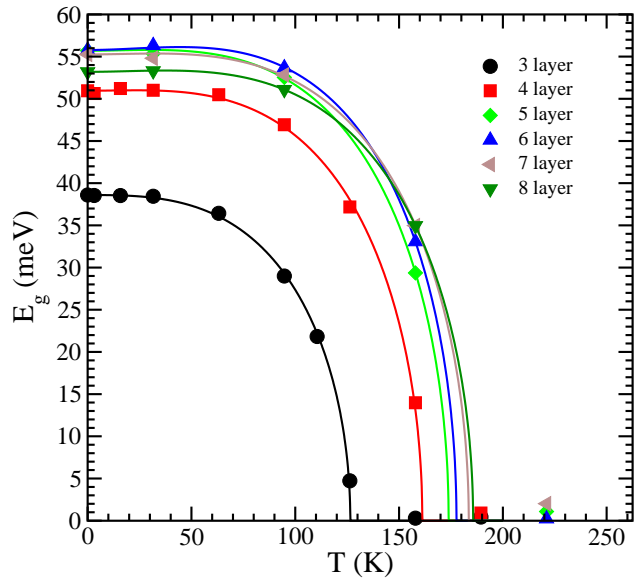}
        \caption{Temperature dependence of the band gap, $E_g$, up to eight layers.
        The dots are the data obtained with PBE0 functional.
        The lines are the result of the fit to eq. (\ref{eq:Tc}).}
        \label{fig:Tc}
\end{figure}

The change in the band gap at zero temperature for different layers
is shown in Figs. \ref{fig:magneticBands} and \ref{fig:Tc} and Table \ref{table:fit}.
The band gap increases significantly from three to four layers, 
but saturates after five layers.
At eight layers, the gap starts to close slowly, with increasing deviation from the flat bands,
in order to reach  the bulk limit \cite{Lin2016} shown in Fig. \ref{fig:BulkoverS}.

As the gap is strongly temperature dependent in experiments and closes
at $T_c=34$ K\cite{Guinea2014}, we study
the antiferromagnetic gap as a function of temperature. This is done
by including a Fermi-Dirac occupation of the electronic states.
The temperature dependence of the gap is shown in 
Fig. \ref{fig:Tc}. To obtain  $T_c$, 
we fit the band gap to the function,
\begin{eqnarray}
E_g(T)=E_g(0) && \left[  A\left(1-\frac{T}{T_c}\right) +(3-2A)\left(1-\frac{T}{T_c}\right)^2 \right. \nonumber \\ 
                          && \left. +(A-2)\left(1-\frac{T}{T_c}\right)^3 \right]^{1/2} 
\label{eq:Tc}                          
\end{eqnarray}
with the constants arranged such that 
the first and the second derivative of the curve is zero
at the zero temperature limit.
The temperature dependence of the band gap has a similar behavior to 
that obtained from the experiments in Ref. \cite{Guinea2014}.
However, the  thermal suppression of the gap is slower
in our calculations, resulting in a larger $T_c=126.5$ K,
as compared to the experimental value of $T_c=34$ K for trilayer
graphene.
We attribute the discrepancy between theory and experiments 
to the imperfect treatment of screening at the hybrid functional level. 

Similarly to what happens for the layer dependence of the band gap with the number of layers,
$T_c$ increases up to five layers, and then saturates and changes slowly.
The $T_c$ values and the fit parameter $A$ of each layer are shown in Table \ref{table:fit}
for different numbers of layers.
\begin{table}[!htb] \footnotesize
\caption{The $T_c$ values and the fit parameter $A$ of eq. (\ref{eq:Tc}) of each layer.
		}
\centering
\begin{ruledtabular}
        \begin{tabular}{c c c c}
        N & $E_g(0)$ (meV) & $T_c$ (K) & $A$ \\
        \hline
        3 & 38.60 & 126.53 & 2.97 \\
        4 & 50.96 & 161.21 & 3.28 \\
        5 & 55.70 & 173.90 & 3.36 \\
        6 & 55.78 & 177.67 & 3.60 \\
        7 & 55.26 & 183.58 & 3.36 \\
        8 & 53.19 & 185.67 & 3.43 \\
        \end{tabular}
        \end{ruledtabular}
\label{table:fit}
\end{table}

Contrary to the case of suspended samples, supported RSMG shows a
metallic behavior \cite{Thomas2015}. A possible reason for this
could be the presence of a substrate doping \cite{Guinea2014}.
In order to verify this
hypothesis, we consider n-doping of rhombohedral trilayer graphene. We use a
compensating jellium background to enforce charge neutrality.

To estimate the number of electrons needed to close the gap,
we have integrated the first peak of the density of states
in Fig. \ref{fig:magneticBands}.
For the trilayer rhombohedral graphene an electron density
of $n=8.01 \times 10^{11}$ cm$^{-2}$
(corresponding to $x=0.00042$ electrons/cell)
is needed to fill the flat conduction band region.
This electron density 
increases to $n=67.70 \times 10^{11}$ cm$^{-2}$ for eight layers. 
The results for the rest of the layers are presented in Appendix \ref{elecFlatBand}.

We find that doping reduces the band gap. 
In agreement with our estimation from the density of states,
the gap ultimately closes at 
$n \sim 7 \times 10^{11}$ cm$^{-2}$, to be compared to 
 $n \sim 3 \times 10^{11} $ cm$^{-2}$ found in
 experiments \cite{Guinea2014}.
The combined effect of doping and temperature is shown in Fig. \ref{fig:ndop} 
(and in Appendix \ref{GapTDop}).
We conclude that substrate doping can be responsible of the
differences between supported and suspended samples.
We can also speculate that the critical doping to close the gap could grow
by an order of magnitude by increasing the number of ABC stacked layers.

\begin{figure}[!htb]
        \centering
                \includegraphics[clip=true, trim=0mm 55mm -3mm -55mm, width=0.48\textwidth]{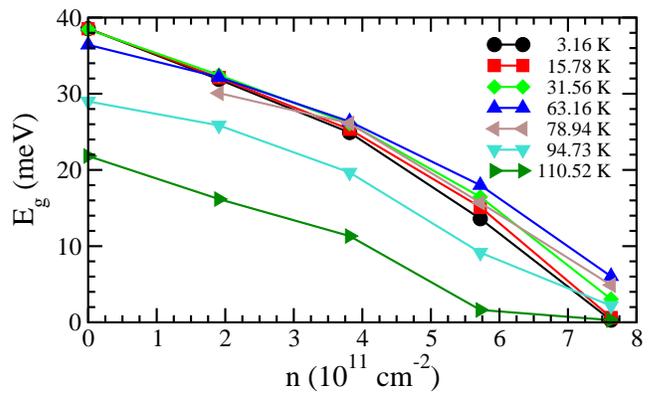}
        \caption{Doping dependence of the band gap of rhombohedral trilayer graphene
        at different temperatures.
        The lines are to guide the eye.}
        \label{fig:ndop}
\end{figure}

Up to now we consider the possibility of magnetic gap opening in
RSMG. However, charge density wave instabilities could also open a gap
via phonon softening. In order to validate this hypothesis, we
calculate phonon frequencies at ${\mathbf \Gamma}$ and ${\bf K}$ by using
the finite differences method and the spinless PBE0 functional
in rhombohedral-stacked trilayer graphene. 
We find that the largest softening occurs
at the \textbf{K} point of the Brillouin zone
\cite{Attaccalite2008} for the in-plane phonon mode.
Despite PBE0
substantially softens the phonon frequencies with respect to PBE, 
all the phonon modes are stable so that a structural distortion
compatible with a $\sqrt{3}\times\sqrt{3}$ periodicity is excluded. 
The magnitude of the phonon modes are also reported in Appendix \ref{Phon}.
Magnetism is then the most likely instability in RSMG.

In this paper, we have analyzed the magnetic and charge
instabilities in rhombohedral
stacked multilayer graphene using spin-polarized hybrid functionals.
While in the absence of spin-polarization an extremely flat
surface state occurs at the Fermi level, the introduction of spin
polarization leads to magnetic instabilities and opening of a gap in
the surface state. 
The state is such that the surface layers are weakly ferrimagnetic in-plane,
and the top and the bottom layers are antiferromagnetic coupled,
which is propagated by the out-of-plane antiferromagnetic coupling between the 
nearest neighboring atoms.
The globally stable state is antiferromagnetic where the spin up and spin down 
bands are degenerate.
Within PBE0 the gap is found to agree with experiments on ABC trilayer graphene.
We have shown that doping suppress the gap, explaining the
experimental finding that a gap occurs in trilayer ABC graphene only
in suspended samples.
Finally, we study the possible occurrence of charge density wave
instabilities with $\sqrt{3}\times\sqrt{3}$ supercell. We found
that no charge density wave occurs so that the gap opening seen in 
experiments is only due to the magnetic coupling of the surface atoms
in the multilayers.
Our work demonstrate that the inclusion of exact exchange 
in first principles calculations and an ultradense sampling 
of the Brillouin zone are crucial in order to explain the magnetic and structural
instabilities of rhombohedral-stacked multilayer graphene.

\begin{acknowledgments}
This work is supported by the Graphene Flagship, and by Prace (Proposal number 2015133134). 
Calculations were performed at IDRIS, CINES, CEA and BSC TGCC.
\end{acknowledgments}

\appendix
\section{Exchange and Correlation Functionals}
\label{XCfcnal}

\begin{figure}[!htb]
        \centering
                \includegraphics[clip=true, trim=-5mm 10mm 0mm -15mm, width=0.48\textwidth]{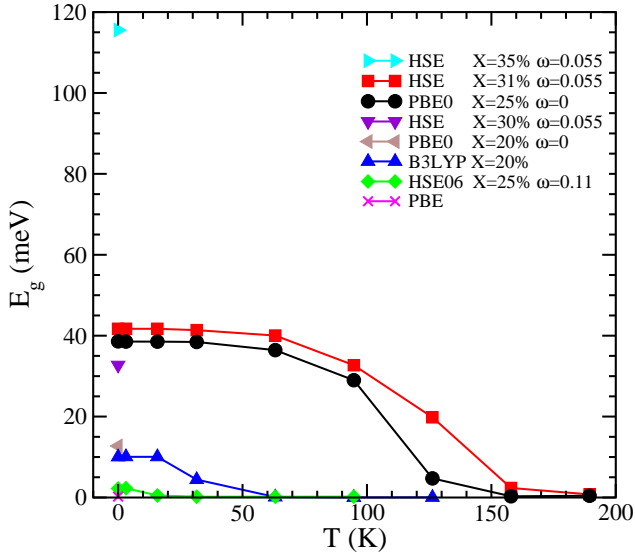}
        \caption{Change in the band gap with different DFT functionals 
        with different exact exchange (X) percentage (\%) and range separation ($\omega$) in \AA$^{-1}$
        for rhombohedral trilayer graphene.
        The lines are to guide the eye.}
        \label{fig:EgXC}
\end{figure}

To understand the effect of exchange and range separation in the hybrid functionals
on the band gap of the system, we have tuned exchange components and range separation.
The results are given in Fig. \ref{fig:EgXC}.

For the same range separation parameter, $\omega=0.055$ \AA$^{-1}$,
the increasing percentage of exact exchange increases the band gap significantly.
For the same percentage of exact exchange, increasing the range separation
decreases the band gap, as can be seen when comparing the 
PBE0 functional to the HSE06 functional \cite{HSE06}.

Only with PBE0 functional, we obtain a band gap similar to the experimental value.

When we change the exact exchange and range separation of the HSE functional
such that the band gap is similar to that of PBE0 at zero temperature limit
(X$=31\%, \omega=0.055$ \AA$^{-1}$),
we obtain a similar temperature dependence for both functionals.

With the B3LYP functional \cite{b3lyp}, we obtain a $T_c$ similar to the experimental value,
however the calculated band gap at zero temperature is too small compared to the
experimental result \cite{Guinea2014}.

\begin{table}[!htb] \footnotesize
\caption{The magnitude of the spin of each atom in $10^{-3}\mu_B$ at $T=0$ K for
		rhombohedral trilayer graphene calculated by
		different exchange and correlation functionals (XC) with different 
		exact exchange (X) percentage (\%) and range separation ($\omega$) in \AA$^{-1}$.
		}
\centering
\begin{ruledtabular}
        \begin{tabular}{l r l r r r}
		XC & X & $\omega$ & $\mu_1=-\mu_6$ & $\mu_2=-\mu_5$ & $\mu_3=-\mu_4$ \\
        \hline
        PBE   &  0 & 0     &  -0.0001 &  0.0001 &  -0.0002 \\
        HSE06 & 25 & 0.11  &  -0.4589 &  0.3581 &  -0.1282 \\
        B3LYP & 20 & 0     &  -1.4442 &  1.1316 &  -0.5043 \\
        PBE0  & 20 & 0     &  -1.8892 &  1.5258 &  -0.7099 \\
        HSE   & 30 & 0.055 &  -5.2116 &  4.5689 &  -2.6368 \\
        PBE0  & 25 & 0     &  -5.2831 &  4.5951 &  -2.7317 \\
        HSE   & 31 & 0.055 &  -6.5798 &  5.8432 &  -3.5352 \\
        HSE   & 35 & 0.055 & -17.5428 & 16.2741 & -12.3476 \\
        \end{tabular}
        \end{ruledtabular}
\label{table:XCspin}
\end{table}

The value of the band gap is directly linked to the magnitude of the spin in the surface
atoms, which can be seen in Table \ref{table:XCspin}.
The PBE functional predicts a spinless paramagnetic state, 
while the introduction of the exact exchange
immediately stabilizes the magnetic state.
With the increase of the amount of exact exchange the magnitude of the spin
on the surface atoms increases,
and with the increase of the range separation the magnitude of the spin
of the surface atoms decreases. 

\section{Projected Band Structure}
\label{PBand}

\begin{figure}[!thb]
        \centering
                \includegraphics[clip=true, trim=-20mm 15mm 30mm 10mm, width=0.48\textwidth]{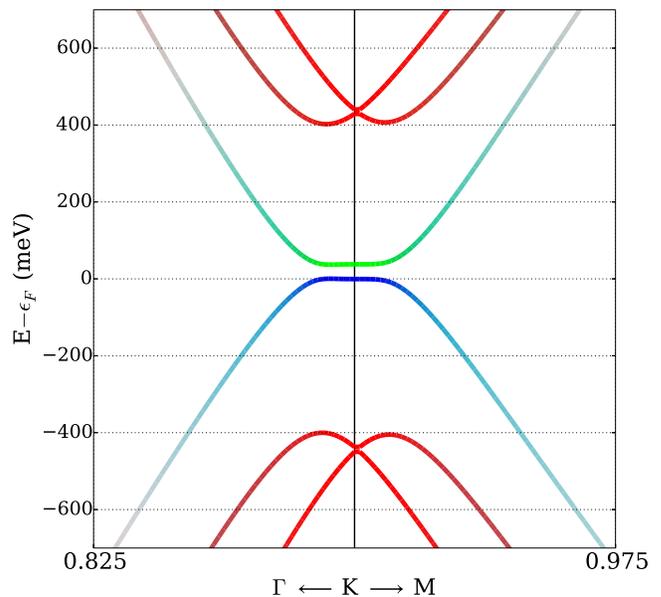}
        \caption{The electronic band structure of trilayer rhombohedral graphene
        projected onto $p_z$ orbital spin-down state
        of atom 1 (blue), 6 (green), and 2 and 5 combined (red).
        The electronic bands are plotted around 0.075 bohr$^{-1}$ of the \textbf{K} point
        along the path $\mathbf{\Gamma} \rightarrow \mathbf{K} \rightarrow \mathbf{M}$.}
        \label{fig:pBands}
\end{figure}

In Fig. \ref{fig:pBands}, we present the electronic band structure 
projected onto the spin-down state of $p_z$ orbital.
The flat bands are dominated by atom 1 (blue in the figure) and 6 (green),
for the spin down bands.
These are one of the atoms on each surface, and are antiferromagnetic coupled.
The other atom of each surface, i.e., atoms 2 and 5 (red), contribute only to the bulk bands.
The spin-up figure would look exactly the same except
now blue would be atom 6 and green would be atom 1.

In order to understand the interplay between in-plane and out-of-plane
magnetic couplings in determining the gap structure we performed,
for three and four layers,
a calculation starting from an in-plane antiferromagnetic spin order
and an out-of-plane ferromagnetic order.
The self-consistent cycle preserves this magnetic state; however,
the resulting band structure is gapless.
This reveals that the inter-layer antiferromagnetic coupling plays a crucial role
in the gap opening in three- and four-layer rhombohedral graphene.

\section{Metallic and Paramagnetic Bands}
\label{MetalBand}

\begin{figure}[!thb]
        \centering
                \includegraphics[clip=true, trim=0mm 0mm -10mm -5mm, width=0.48\textwidth]{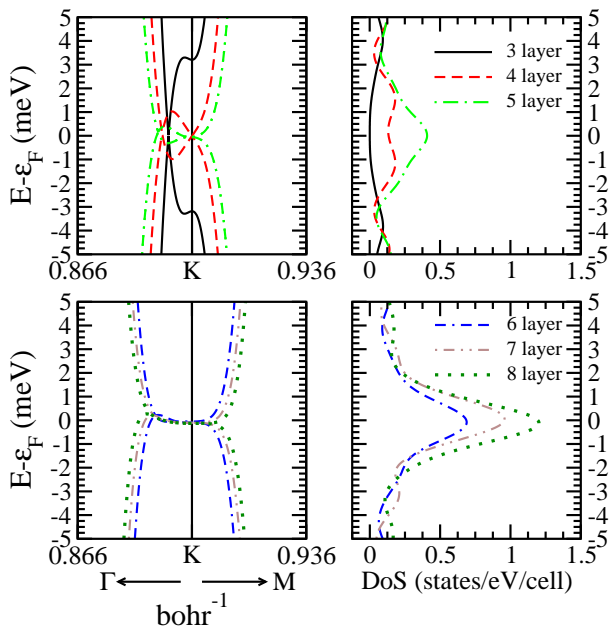}
        \caption{The electronic band structure (left) and density of states (right)
        of the surface states of the paramagnetic state of the rhombohedral graphene 
        up to eight layers.
        The electronic bands are plotted around 0.035 bohr$^{-1}$ of the \textbf{K} point
        along the path $\mathbf{\Gamma} \rightarrow \mathbf{K} \rightarrow \mathbf{M}$.}
        \label{fig:paramagBands}
\end{figure}

In Fig. \ref{fig:paramagBands}, we present the electronic band structure 
and the density of states of the paramagnetic state
calculated with the PBE0 functional between three and eight layers.

The paramagnetic electronic band calculations are performed with the same parameters 
as the magnetic calculations,
except the initial conditions on the spin of each atom are not set
and Fermi-Dirac smearing of 0.00001 Ha is used.
The density of states is calculated with a Gaussian smearing of 0.00004 Ha.

The crossing points of the bands in our paramagnetic calculations are
comparable to the previous results obtained for three and four layers with 
standard DFT calculations \cite{Henrard2006}.

As compared to the bulk bands of rhombohedral graphite with DFT \cite{Lin2011,Michenaud1994,Shyu2011,Taut2011}
and tight binding \cite{Shyu2011} calculations,
and to the evolution of graphene to graphite with tight binding calculation \cite{Lin2016},
it is clear that these states are surface states of the 
few-layer rhombohedral graphene.

\section{Electrons on the Flat Surface Bands}
\label{elecFlatBand}

In order to understand the amount of charge needed to fill the flat surface band
and close the gap,
we have integrated first peak of the density of states above the gap.
The results for each layer are shown in Table \ref{table:ndoplayer}.

\begin{table}[!htb] \footnotesize
\caption{The number of electrons, $x$, in units of electrons/cell
		and the electron density, $n$, in units of $10^{11}$ cm$^{-2}$,
		needed to fill the flat surface bands, for each layer.
		}
\centering
\begin{ruledtabular}
        \begin{tabular}{c c c}
		N & $x$ & $n$ \\
        \hline
		3 & 0.00042 & ~8.01 \\ 
		4 & 0.00116 & 22.12 \\ 
		5 & 0.00209 & 39.86 \\ 
		6 & 0.00268 & 47.30 \\ 
		7 & 0.00326 & 62.17 \\ 
		8 & 0.00355 & 67.70 \\ 
        \end{tabular}
        \end{ruledtabular}
\label{table:ndoplayer}
\end{table}

For the trilayer rhombohedral graphene, a doping of $x=0.00042$ electrons/cell 
is needed to fill the flat conduction band,
and this is in agreement with our calculations that at $\sim 0.0004$ electrons/cell
the band gap closes, as will be discussed in the following Appendix.
Note that the width of the flat region decreases with increasing doping.

\section{Band Gap with Changing Temperature and Doping}
\label{GapTDop}

\begin{figure}[!thb]
        \centering
                \includegraphics[clip=true, trim=0mm 10mm -5mm -15mm, width=0.48\textwidth]{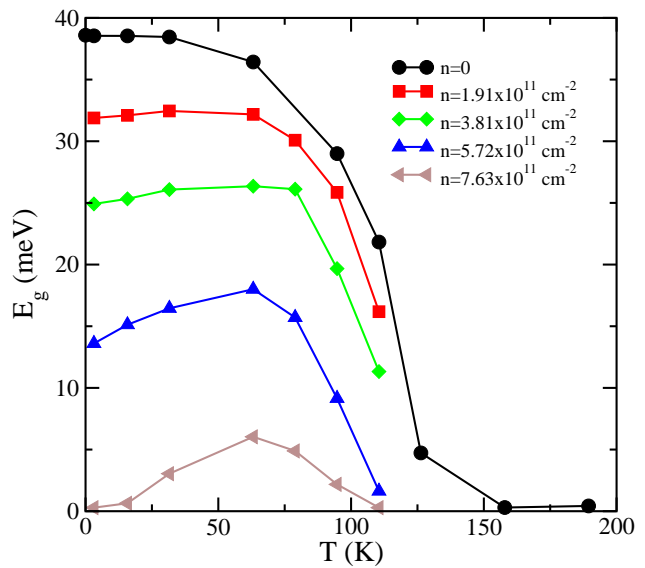}
        \caption{Temperature dependence of the band gap for rhombohedral trilayer graphene
        with different doping.
        The dots are the data obtained with PBE0 functional.
        The lines are to guide the eye.}
        \label{fig:Tcndop}
\end{figure}

\begin{figure*}[!hbt]
        \centering
                \includegraphics[clip=true, trim=40mm 18mm -40mm -30mm, width=1\textwidth]{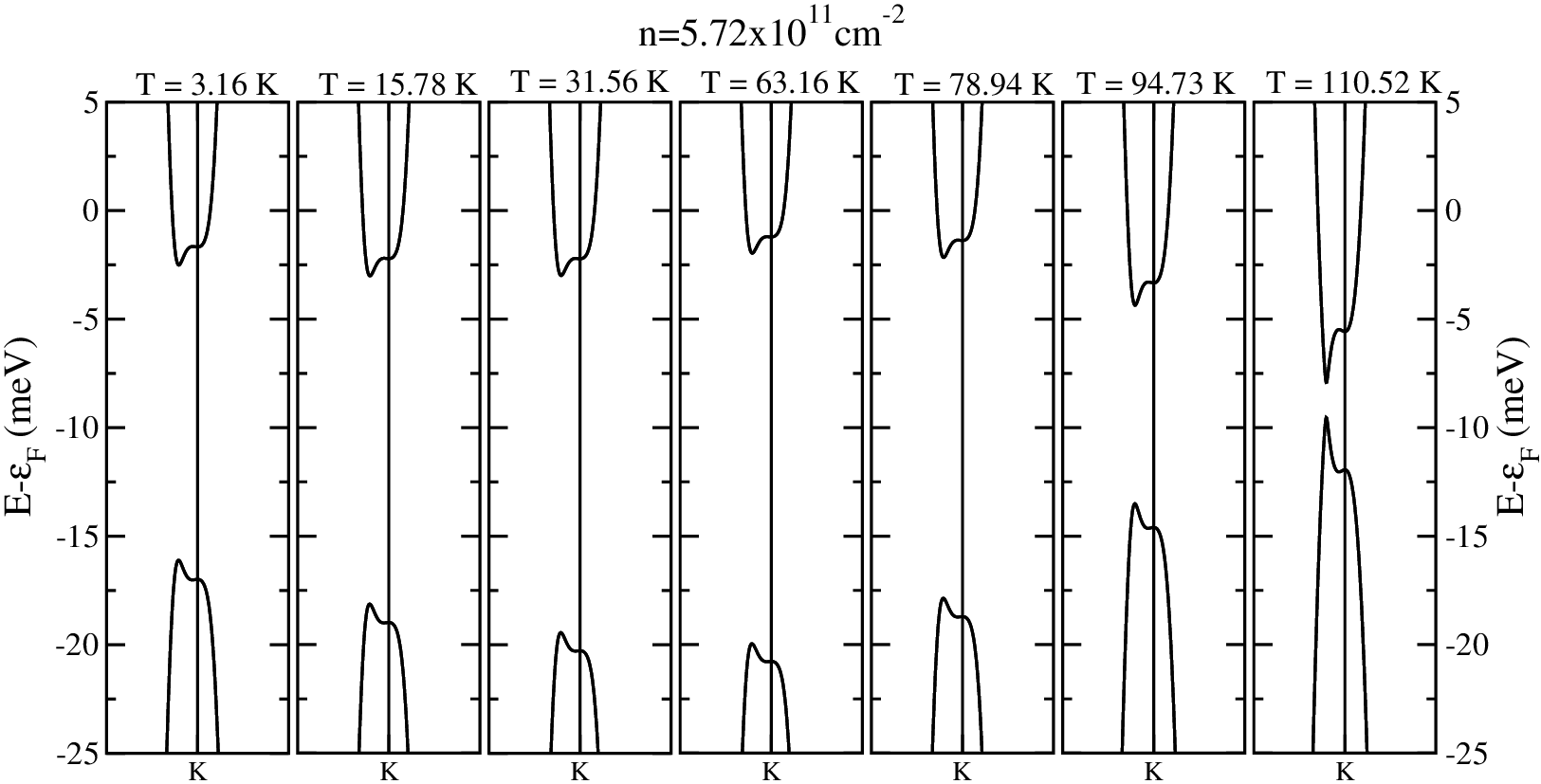}
        \caption{The electronic band structure 
        of the surface states of the magnetic state of the rhombohedral trilayer graphene 
        at doping $n=5.72\times10^{11}$ cm$^{-2}$ for different temperatures.
        The electronic bands are plotted around 0.035 bohr$^{-1}$ of the \textbf{K} point
        along the path $\mathbf{\Gamma} \rightarrow \mathbf{K} \rightarrow \mathbf{M}$.}
        \label{fig:bandsDop}
\end{figure*}

\begin{figure*}[!hbt]
        \centering
                \includegraphics[clip=true, trim=0mm 4mm -10mm -30mm, width=1\textwidth]{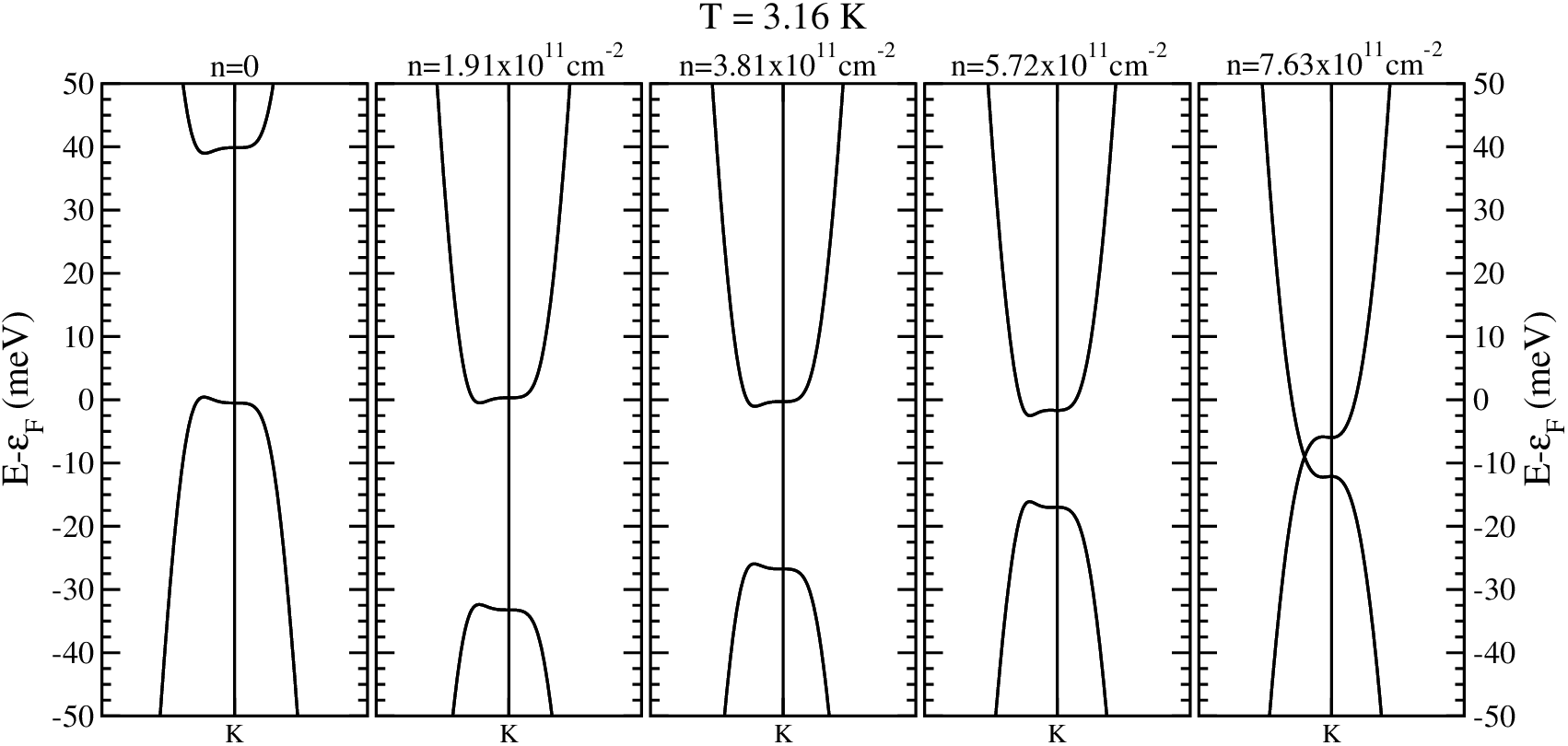}
        \caption{The electronic band structure 
        of the surface states of the magnetic state of the rhombohedral trilayer graphene 
        at FD smearing temperature $T=3.16$ K for different doping.
        The electronic bands are plotted around 0.035 bohr$^{-1}$ of the \textbf{K} point
        along the path $\mathbf{\Gamma} \rightarrow \mathbf{K} \rightarrow \mathbf{M}$.}
        \label{fig:bandsDopT3}
\end{figure*}

\begin{figure*}[!hbt]
        \centering
                \includegraphics[clip=true, trim=0mm 4mm -10mm -30mm, width=1\textwidth]{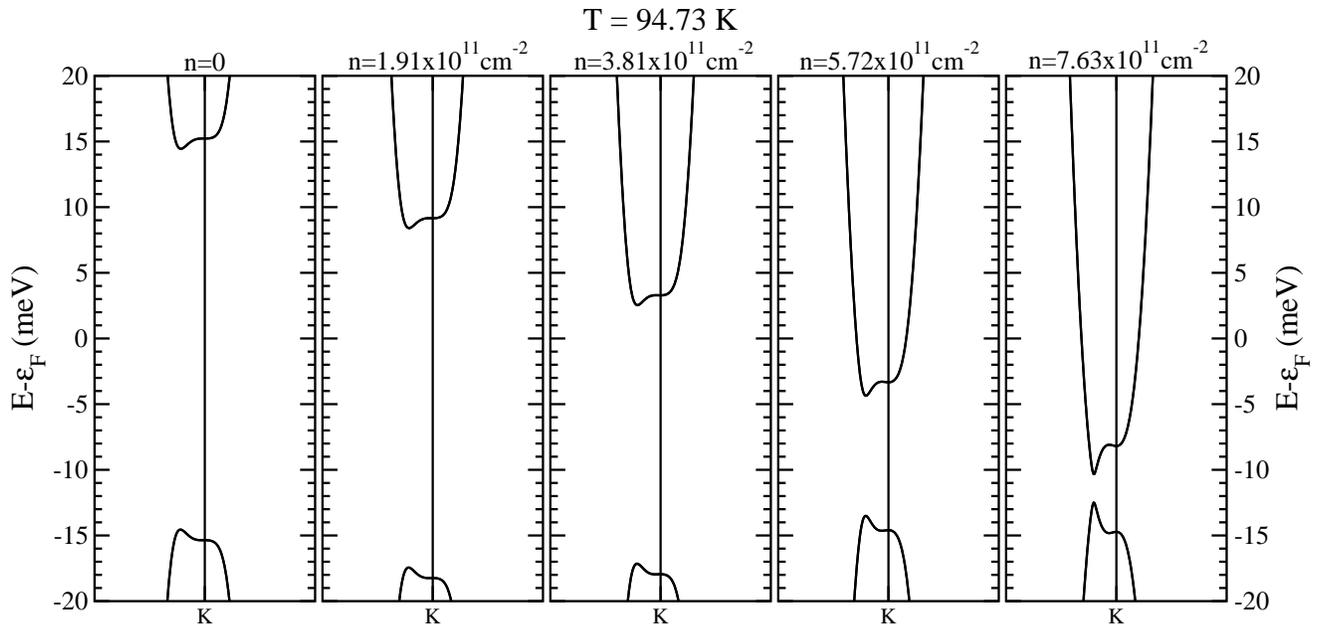}
        \caption{The electronic band structure 
        of the surface states of the magnetic state of the rhombohedral trilayer graphene 
        at FD smearing temperature $T=94.73$ K for different doping.
        The electronic bands are plotted around 0.035 bohr$^{-1}$ of the \textbf{K} point
        along the path $\mathbf{\Gamma} \rightarrow \mathbf{K} \rightarrow \mathbf{M}$.}
        \label{fig:bandsDopT94}
\end{figure*}

To understand the effect of the doping on the band gap, 
we introduced $x=0.0001, 0.0002, 0.0003, 0.0004$ electrons/cell
(corresponding to the electron density of $n=1.91\times10^{11}, 3.81\times10^{11}, 5.72\times10^{11}, 7.63\times10^{11}$ cm$^{-2}$)
to the trilayer rhombohedral graphene.
In Fig. \ref{fig:Tcndop}, we present the temperature dependence of the band gap
for different doping, as compared to the undoped case.

With the increasing temperature, after $T \sim 60$ K the band gap decreases sharply,
as expected.
However, at low temperatures, the band gap first increases.
In Fig. \ref{fig:bandsDop}, we present the electronic structure 
at doping $n=5.72\times10^{11}$ cm$^{-2}$ for different temperatures.
In addition, to understand the correlation between the increase in the gap and the spins,
we present the spins of each atom at different temperatures for this doping
in Table \ref{table:spindope}. 

\begin{table}[!htb] \footnotesize
\caption{The magnitude of the spin of each atom in $10^{-3}\mu_B$ at each temperature
		for doping $n=5.72\times10^{11}$ cm$^{-2}$.
		}
\centering
\begin{ruledtabular}
        \begin{tabular}{l c c c}
		T (K) & $\mu_1=-\mu_6$ & $\mu_2=-\mu_5$ & $\mu_3=-\mu_4$ \\
        \hline
          3.16 & -1.86 & 1.62 & -1.00 \\
         15.78 & -2.06 & 1.79 & -1.10 \\
         31.56 & -2.24 & 1.95 & -1.17 \\
         63.16 & -2.46 & 2.14 & -1.29 \\
         78.94 & -2.14 & 1.86 & -1.12 \\
         94.73 & -1.25 & 1.09 & -0.65 \\
        110.52 & -0.22 & 0.19 & -0.12 \\
        \end{tabular}
        \end{ruledtabular}
\label{table:spindope}
\end{table}

Furthermore, we also present the electronic structure at two different temperatures
$T=3.16$ K and $T=94.73$ K for different doping in Figs. \ref{fig:bandsDopT3} and 
\ref{fig:bandsDopT94}, respectively.
The closing of the band gap with increasing doping is clear from these figures.
Also note that the width of the flat region decreases with increasing doping.

\section{Phonon Modes}
\label{Phon}

\begin{table}[!htb] \footnotesize
\caption{The frequencies in cm$^{-1}$ at the $\mathbf{\Gamma}$ and $\mathbf{K}=\mathbf{K'}$ points of the
		rhombohedral trilayer graphene.}
\centering
\begin{ruledtabular}
        \begin{tabular}{r r}
        $\mathbf{\Gamma}$ & $\mathbf{K}$ \\
        \hline
     0.0000 &  547.8938 \\
     0.0000 &  547.8938 \\
     0.0000 &  549.8043 \\
    19.0449 &  549.8138 \\
    19.0449 &  555.7702 \\
    31.5473 &  555.7702 \\
    31.5473 & 1011.7181 \\
    71.1469 & 1015.4720 \\
   120.6523 & 1015.4720 \\
   762.5849 & 1192.2007 \\
   780.7989 & 1192.2007 \\
   788.6079 & 1211.2703 \\
  1601.4634 & 1249.0268 \\
  1601.4634 & 1249.1592 \\
  1607.1849 & 1249.3243 \\ 
  1613.3836 & 1249.3243 \\ 
  1613.3836 & 1250.3788 \\
        \end{tabular}
        \end{ruledtabular}
\label{table:phon}
\end{table}

We have calculated the phonon modes with the PBE0 functional
using a Fermi-Dirac smearing of 0.002 Ha,
electronic k mesh of $39 \times 39 \times 1$,
energy convergence tolerance of $10^{-9}$ Ha,
real space integration tolerances of 7-7-7-15-30,
and a $\sqrt{3}\times\sqrt{3}\times1$ supercell
to obtain the modes at both the $\mathbf{\Gamma}$ and $\mathbf{K}$ points.

As presented in Table \ref{table:phon}, all the phonon modes 
at the $\mathbf{\Gamma}$ and $\mathbf{K}$ points of the Brillouin zone
are positive.
Therefore, we conclude that there is no charge density wave instability
with $\sqrt{3}\times\sqrt{3}$ periodicity
to cause the opening of the band gap for this system.

The three phonon modes with a large electron-phonon coupling are 
at the $\mathbf{K}$ point: 
two degenerate modes at 1192.2007 cm$^{-1}$ and the other mode at 1211.2703 cm$^{-1}$.
They are softened significantly with the PBE0 functional with exact exchange,
as compared to the local density approximation \cite{Chou2008};
however, this softening is not enough to cause an instability.
Moreover, this softening is the largest compared to other standard hybrid functionals,
since the exact exchange is smaller in the B3LYP and 
range separation is larger in the HSE06 functional.

\bibliography{PaperBetul}

\end{document}